\documentclass[conference, 10pt]{IEEEtran}
\IEEEoverridecommandlockouts
\usepackage{cite}
\usepackage{amsmath,amssymb,amsfonts}
\usepackage{algorithmic}
\usepackage{graphicx}
\usepackage{textcomp}
\usepackage{xcolor}
\usepackage{soul}

\ifCLASSOPTIONcompsoc
    \usepackage[caption=false, font=normalsize, labelfont=sf, textfont=sf]{subfig}
\else
\usepackage[caption=false, font=footnotesize]{subfig}
\fi

\definecolor{mybrown}{RGB}{196,104,54}
\definecolor{mygray}{RGB}{72,86,107}
\definecolor{myyellow}{RGB}{236,204,84}

\usepackage{dblfloatfix}
\usepackage{hyperref}
\usepackage{braket}
\usepackage{listings}
\usepackage{caption}
\usepackage[ruled,vlined, linesnumbered]{algorithm2e}
\SetKwRepeat{Do}{do}{while}%
\definecolor{codegreen}{rgb}{0,0.6,0}
\definecolor{codegray}{rgb}{0.5,0.5,0.5}
\definecolor{codepurple}{rgb}{0.58,0,0.82}
\definecolor{backcolour}{rgb}{0.95,0.95,0.92}

\lstdefinestyle{mystyle}{
  backgroundcolor=\color{backcolour},   commentstyle=\color{codegreen},
  keywordstyle=\color{magenta},
  numberstyle=\tiny\color{codegray},
  stringstyle=\color{codepurple},
  basicstyle=\ttfamily\footnotesize,
  breakatwhitespace=false,         
  breaklines=true,                 
  captionpos=b,                    
  keepspaces=true,                 
  numbers=left,                    
  numbersep=5pt,                  
  showspaces=false,                
  showstringspaces=false,
  showtabs=false,                  
  tabsize=2
}

\newtheorem{example}{Example}[section]

\def\BibTeX{{\rm B\kern-.05em{\sc i\kern-.025em b}\kern-.08em
    T\kern-.1667em\lower.7ex\hbox{E}\kern-.125emX}}
\begin{document}

\title{Shuttle-Exploiting Attacks and Their Defenses \\in Trapped-Ion Quantum Computers
}

\author{\IEEEauthorblockN{Abdullah Ash Saki}
\IEEEauthorblockA{\textit{Dept. of Electrical Engineering} \\
\textit{Pennsylvania State University}\\
University Park, PA \\
axs1251@psu.edu}
\and
\IEEEauthorblockN{Rasit Onur Topaloglu}
\IEEEauthorblockA{
% \textit{dept. name of organization (of Aff.)} \\
\textit{IBM}\\
Hopewell Junction, NY \\
rasit@us.ibm.com}
\and
\IEEEauthorblockN{Swaroop Ghosh}
\IEEEauthorblockA{\textit{Dept. of Electrical Engineering} \\
\textit{Pennsylvania State University}\\
University Park, PA \\
szg212@psu.edu}
}

\maketitle
\thispagestyle{plain}
\pagestyle{plain}

\begin{abstract}
Trapped-ion (TI) quantum bits are a front-runner technology for quantum computing. TI systems with multiple interconnected traps can overcome the hardware connectivity issue inherent in superconducting qubits and can solve practical problems at scale. With a sufficient number of qubits on the horizon, the multi-programming model for Quantum Computers (QC) has been proposed where multiple users share the same QC for their computing. Multi-programming is enticing for quantum cloud providers as it can maximize device utilization, throughput, and profit for clouds. Users can also benefit from the short wait queue. However, shared access to quantum computers can create new security issues. This paper presents one such vulnerability in shared TI systems that require \emph{shuttle} operations for communication among traps. Repeated shuttle operations increase quantum bit energy and degrade the reliability of computations (fidelity). We show adversarial program design approaches requiring numerous shuttles. We propose a random and systematic methodology for adversary program generation. Our analysis shows shuttle-exploiting attacks can substantially degrade the fidelities of victim programs by $\approx 2$X to $\approx 63$X. Finally, we present several countermeasures such as adopting a hybrid initial mapping policy, padding victim programs with dummy qubits, and capping maximum shuttles.
\end{abstract}


\section{Introduction}
Quantum computing has garnered immense attention from industry and academia alike in recent years a quantum computing can potentially speed up certain classes of problems beyond the means of classical computers. It can be advantageous in domains like machine learning~\cite{qml}, drug discovery~\cite{drug}, molecule simulation~\cite{material, kandala}, and optimization~\cite{qaoa}. With an active interest in the field, quantum computing is progressing at a rapid pace. On one end, researchers are proposing new quantum algorithms that leverage unique properties like superposition, entanglement, and interference to speed up computation. On the other end, researchers are pursuing various technologies like superconducting, trapped-ion (TI), and photonics to design quantum bits or qubits. The quantum hardware is also improving over time (i.e., size and quality). 

\begin{figure}
    \centering
    \includegraphics[width=3.5in]{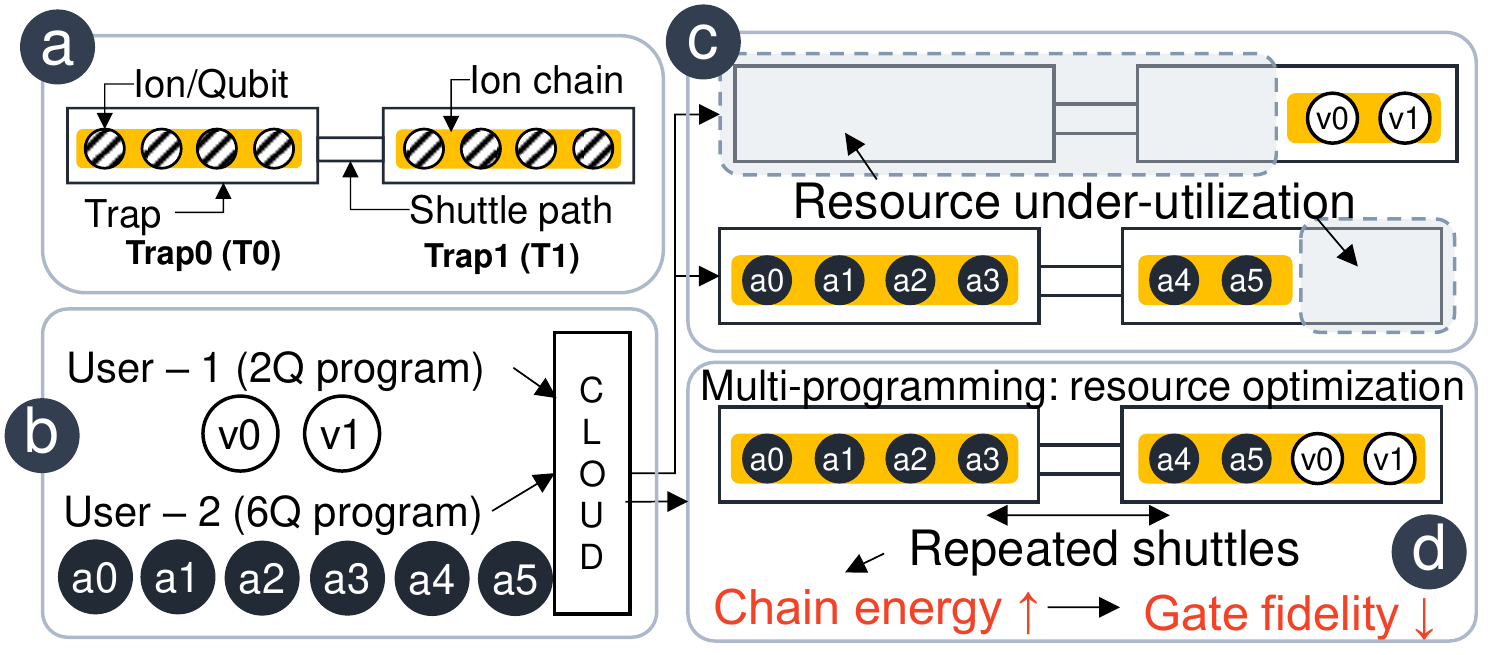}
    \caption{(a) Overview of a TI system. (b) The basic concept of multi-programming. (C) Attack overview: the adversary program shares a trap with the victim. The adversary program tries forcing repeated shuttles between traps which increase chain-energy and degrade gate fidelity in the shared trap. \vspace{-3mm}}
    \label{fig:attack-intro}
\end{figure}

The TI qubit is one of the most promising technologies for building a quantum computer. It offers several advantages such as identical qubits, long coherence times, and all-to-all connectivity among qubits~\cite{ionq}. Several companies such as IonQ and Honeywell are developing TI systems. Recently, Honeywell reported a trapped-ion system with quantum volume (QV) of $512$~\cite{honeywell-qv-512}, highest thus far (QV is a metric proposed by IBM~\cite{qv} to quantify the capabilities of a quantum computer). 
Some of these TI qubit-based systems are also commercially available via IBM Qiskit~\cite{ionq-qiskit, ibm-partners}, Amazon Braket~\cite{aws}, and Microsoft Azure Quantum~\cite{azure-quantum}. Although existing TI systems have a smaller number of qubits compared to their superconducting counterparts (IBM Quantum Eagle processor has $127$ qubits whereas largest known trapped-ion system has $11$ qubits~\cite{ionq}), road-maps for larger systems with $50$-$100$ qubits are in place~\cite{honeywell-h5, staq}. Confining many ions in a single trap becomes problematic from a control and gate implementation perspective. Therefore, the pathway to scalability in TI systems involves multiple interconnected traps. A technology named quantum charge-coupled device (QCCD) is proposed in~\cite{qccd} for scalable and modular trapped-ion systems. In this light, Murali et al.~\cite{murali-ti} performed extensive architectural studies for multi-trap trapped ion systems. They developed a compiler and a simulator~\cite{qccd-github} for such systems with experimentally calibrated values. With industrial and academic efforts, the future of the TI system looks promising.

Fig.~\ref{fig:attack-intro}a shows a TI quantum computer diagrammatically. In a TI system, qubits are realized using ions. Data is encoded as ion's internal states. Ions are confined inside traps using direct current and oscillatory potentials (thus, we use the terms ions and qubits interchangeably for TI systems in this paper).  Fig.~\ref{fig:attack-intro}a shows a $2$-trap system interconnected by a \emph{shuttle path} that allows movement (shuttle) of ions between traps. Here, we are assuming that each trap can accommodate a maximum of $4$ ions i.e., trap capacity = $4$ per trap for illustration purpose only. Note, the work in~\cite{murali-ti} proposed a trap capacity between $15$-$25$ qubits for practical systems and we use this range in our analysis as well. Ions are first cooled and initialized. Then, laser pulses are applied in sequence on the ions to manipulate ions' states to perform computation (quantum gates). Sometimes computation is required on data from ions in different traps. In such cases, one ion is shuttled (moved) from one trap to another so that the ions are co-located, and the gate can be performed. Finally, light is shined on ions and the data is measured as either $0$ or $1$ based on presence or absence of fluorescence.

Besides the computational aspects, \emph{security} is an equally pivotal aspect for any computing paradigm. Several academic studies identifying security issues and fixes are surfacing in the quantum domain~\cite{samah-acm, samah-iccad, samah-ets, qupuf, xtalk-saki, aks-obfus}. In~\cite{samah-iccad}, authors present an attack model where a rogue element in the cloud can report inaccurate device calibration data. Based on the incorrect data, a user may run his/her program on inferior qubits leading to poor results. They propose to include checkpoint circuits in the program to indicate unwanted variation in device calibration data. The attack model in~\cite{qupuf} assumes an untrusted element in the cloud could schedule a program on inferior hardware instead of allocating the requested hardware. They propose a quantum physical unclonable function (QuPUF) to authenticate the requested hardware. Depending on the noise characteristics, each quantum computer demonstrates a unique signature. The authors in~\cite{qupuf} leverage such signature to design the QuPUF and differentiate among the hardware.

A new type of attack vector can emerge in the multi-programming~\cite{poulami-mp} setting (Fig.~\ref{fig:attack-intro}) for quantum computers. Suppose, two users are submitting their programs to the quantum cloud. User--1 program has $2$ qubits, and user--2 program has $6$ qubits. The cloud can schedule the programs individually on hardware. In such cases, device resources will be underutilized. For the $2$-qubit program $6$ qubits will be unused, and vice-versa (i.e., 2 unused qubits for the $6$-qubit program). Multi-programming taps into this gap and proposes scheduling multiple programs together in the same hardware to maximize resource utilization (Fig.~\ref{fig:attack-intro}d). The concept of multi-programming finds its application in commercial quantum clouds such as Rigetti's Quantum Cloud Service (QCS)~\cite{rigetti-qcs} where a user can reserve a \emph{lattice}~\cite{rigetti-qcs-res}. A lattice can be a partial set of qubits from a larger device. Thus, multiple users can run their programs on a different set of qubits from a larger device. Although running multiple programs can optimize resource usage, throughput, and profit for the cloud, it can create security issues. One such security vulnerability in a multi-programming environment is reported in~\cite{xtalk-saki} for superconducting qubits. The authors demonstrate a crosstalk-induced fault injection attack where \emph{crosstalk} from the adversary program affects a victim program. 
However, this attack is not applicable for TI systems due to negligible crosstalk~\cite{qccd-honeywell}.  

In this paper, we present an attack in the multi-programming setting for TI systems by exploiting a new vulnerability in terms of shuttle operations.
Fig.~\ref{fig:attack-intro} provides an overview of the proposed attack model. We assume qubits from the adversary program span over two traps, and they share a trap with qubits with the victim program. For example, adversary and victim qubits share Trap--$1$ (T$1$). The adversary can design his/her program such that it requires computation (gate) between ions from different traps that will need frequent shuttles between traps. Repeated shuttling adds energy to an ion and increases an ion-chain's energy. This elevated chain-energy degrades the reliability of computation (known as \emph{gate fidelity}). As victim qubits share a chain (ion-chain in T$1$) with the adversary qubits, they also suffer from this shuttle-induced fidelity degradation. Although the premise seems simple, there are architectural policies that curbs shuttling and make the attack challenging. Thus, the attack culminates into designing a program that will trick the architectural policies and enforce repeated shuttles. The attack can be launched in a \emph{white-box} setup where the attacker knows the policies and beats them to achieve repeated shuttles (Section~\ref{sec:systematic}). Or, it can be a \emph{black-box} type attack where no prior information is known (Section~\ref{sec:random}).

With the rapid advancement of quantum hardware, quantum algorithm, and quantum architectures, quantum cloud services will become more practical and popular. Now is an opportune time to identify vulnerabilities in and devise appropriate defenses for imminent multi-programming quantum clouds to prepare for practical scale cloud deployments. This paper is one of the first efforts towards this goal.

We make the following contributions in this paper:
\begin{itemize}
    \item Identify repeated shuttle operations as a mode of attack.
    
    \item Present two malicious program generation methodologies - systematic and random. Systematic attack uses prior knowledge about architectural policies to design a strong attack. The random attack does not require any prior knowledge, albeit losing some attack potency.
    
    \item Modify the QCCD-Compiler~\cite{murali-ti, qccd-github} to accommodate multi-programming. We use QCCD-simulator and QCCD-compiler interchangeably in this paper since the compiler and the simulator are a part of the same software tool-chain.  
    
    \item Analyze the impact of trap capacity and victim size~\footnote{We use the following definitions for \emph{program size} and \emph{program length}: program size is the number of qubits in a program and program length is the number of $2$-qubit gates in the program} 
    on shuttle number and fidelity reduction of the victim.
    
    \item Discuss three countermeasures to thwart attacks.
\end{itemize}

The outline of the paper is as follows: Section~\ref{sec:basics} discusses the basics of quantum computing and TI systems. Section~\ref{sec:attack-setup} describes the attack model and the simulation setup for analyses in this paper. Section~\ref{sec:systematic} delineates the methodology of systematic malicious program generation. Section~\ref{sec:random} discusses the random attack program designing principles.  Section~\ref{sec:results} reports the results and discussions. Section~\ref{sec:countermeasures} presents several countermeasures. Finally, Section~\ref{sec:conclusion} draws conclusion.

\section{Basics}\label{sec:basics}
In this section, we discuss the basics of trapped-ion quantum computers and terminologies used in the paper.

\subsection{Qubit and quantum gate}
\subsubsection{Qubits} 
Quantum bits or qubits are the building block of a quantum computer. Qubits store data (i.e., $\ket{0}$ and $\ket{1}$) as various internal states. A qubit can be in both $\ket{0}$ and $\ket{1}$ simultaneously due to quantum superposition property. A qubit state is represented as $\ket{\psi} = a \ket{0} + b \ket{1}$ where $a$ and $b$ are probabilities amplitudes. Measuring the qubit will collapse one state and return classical bits $0$ or $1$ with probabilities $|a|^2$ and $|b|^2$, respectively. 

\subsubsection{Quantum gates}
\begin{figure}
    \centering
    \includegraphics[width=3.2in]{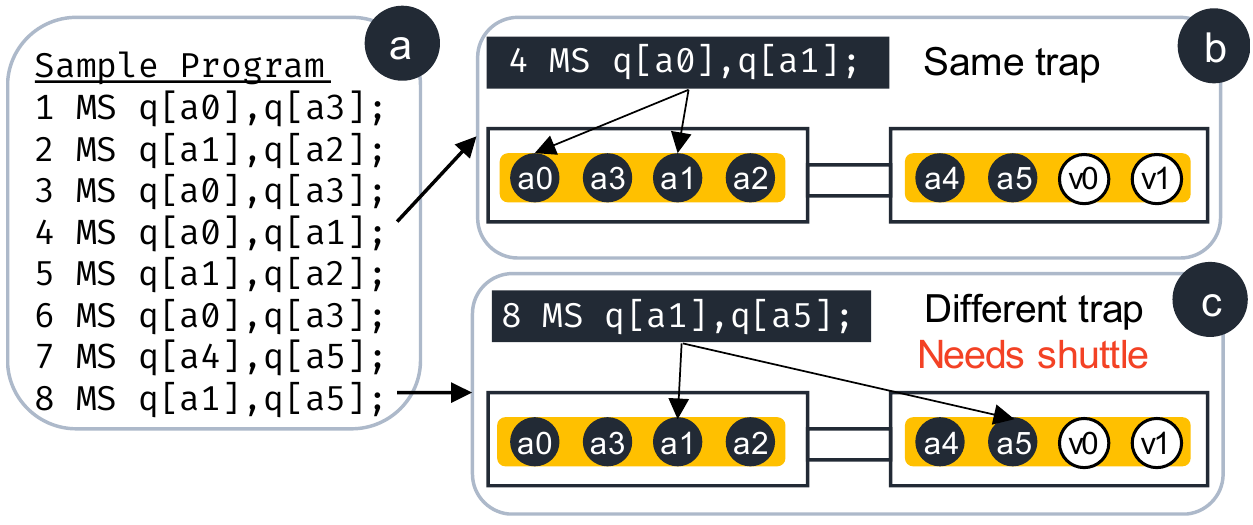}
    \caption{(a) A sample quantum program consisting of $2$-qubit \texttt{MS} gates. We primarily focus on the $2$-qubit gates as it has a lower fidelity and requires shuttle. (b) Ions in same trap: gate can be directly executed. (c) Ions in separate traps: one ion needs shuttling before execution. \vspace{-3mm}}
    \label{fig:need-for-shuttle}
\end{figure}

Quantum gates manipulate information stored in qubits to perform computation. Quantum gates are realized using pulses such as radio frequency (RF) and laser pulses. Gate pulses modify the probability amplitudes of a qubit. For example, a quantum NOT (X) gate pulse when applied to a qubit at state $\ket{0}$ will change amplitudes $a = 1$ and $b = 0$ to $a = 0$ and $b = 1$. Quantum gates are reversible in nature and represented by unitary matrices mathematically. At present, the physically realized gates are $1$-qubit and $2$-qubit. A quantum program is a sequence of quantum gates. Fig.~\ref{fig:need-for-shuttle}a shows a sample quantum program consisting of 2-qubit Mølmer–Sørensen (MS) gates~\cite{ms}.

\subsubsection{Gate fidelity}
Quantum gates in existing quantum computers are erroneous. They incur a finite and non-negligible error rate ($\epsilon$) when executed. Suppose the aforementioned X gate is applied on state $\ket{0}$ for $10,000$ times. One would, in theory, get output $1$ all the time. However, due to gate errors, the user may end up with $9,900$ 1's (correct) and $100$ 0's (incorrect). The gate fidelity ($F$) is usually defined as the complement of the error rate i.e., $F = 1 - \epsilon$. It depends on variations of control pulses and environmental interference. Qubits are kept and operated in a controlled environment to shield them from various noises so that gate fidelities can be high. A lower gate fidelity will introduce more errors in the output and can completely decimate the result.

\begin{figure}
    \centering
    \includegraphics[width=3.2in]{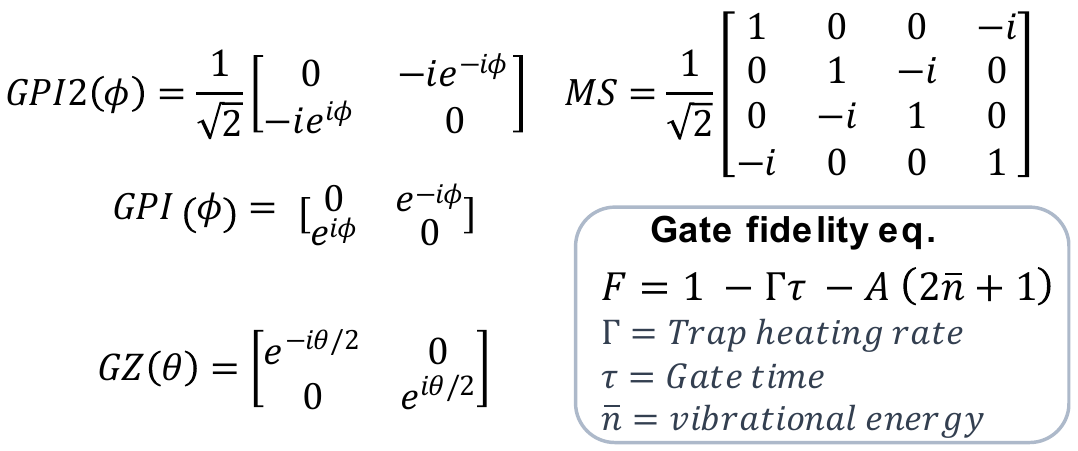}
    \caption{Matrices of gates used in IonQ trapped-ion quantum computer, and the gate fidelity equation presented in~\cite{murali-ti,am1}.}
    \label{fig:gates}
\end{figure}

\begin{figure}
    \centering
    \includegraphics[width=2.5in]{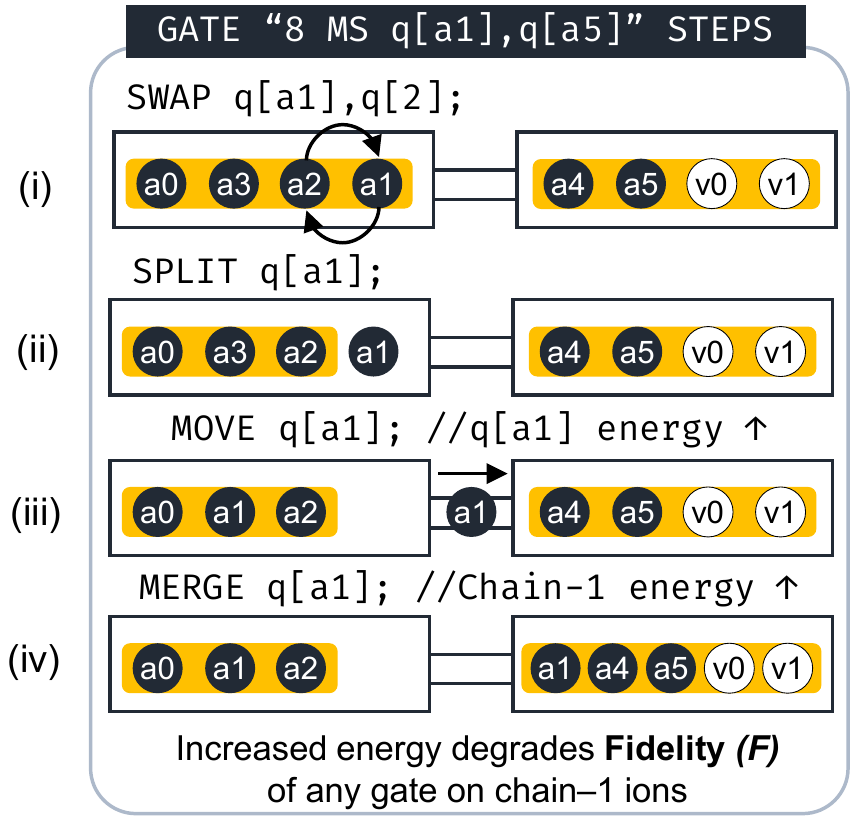}
    \caption{Shuttle steps to bring ions a$1$ and a$5$ in the same trap. \vspace{-6mm}}
    \label{fig:Shuttle steps}
\end{figure}

\subsection{Trapped-ion QC}

\subsubsection{Trap details}
In a trapped-ion system, atoms like Yb or Ca are ionized and trapped between electrodes using electromagnetic fields. Hence, the name trapped-ion quantum computer. Data $\ket{0}$ and $\ket{1}$ are encoded as internal states such as hyper-fine or Zeeman states of the ions. Fig.~\ref{fig:attack-intro} shows the schematic of a trapped-ion system. It has 2 traps: Trap $0$ or T$0$ and Trap $1$ or T$1$. Inside the traps, ions form chains. The traps are connected by a shuttle path which allows movement (shuttle) of an ion from one trap to another if needed. Traps can accommodate a certain number of ions known as \emph{trap capacity}. For example, traps in Fig.~\ref{fig:attack-intro} have a trap capacity of $4$ ions per trap. Besides, some capacity is reserved for incoming ions from other traps known as \emph{communication capacity}. The communication capacity is not shown explicitly in Fig.~\ref{fig:attack-intro}, however, an ion can be moved from T$0$ to T$1$, and in that case, T$1$ will hold $5$ ions. Thus, the $trap~capacity~+~communication~capacity$ defines the absolute maximum number of ions a trap can hold. communication capacity is much smaller than trap capacity in general~\cite{murali-ti}. Finally, the excess cap (EC) of a trap is defined as \emph{trap capacity + communication capacity - ions in the trap}.

\subsubsection{Gate details}
Laser pulses are used to perform quantum gate operations on the qubits/ions. Mølmer–Sørensen (MS) gate is the typical native $2$-qubit gate of trapped-ion systems~\cite{ms}. It is accompanied by several $1$-qubit gates which are mainly rotation gates to form a universal gate-set. For example, the $1$-qubit gates are GPI, GPI2, and GZ in the IonQ system~\cite{ionq-gates}. The matrices for these gates are shown in Fig.~\ref{fig:gates}. 
Inside a trap, all the qubits are connected meaning a $2$-qubit gate can be performed between any two qubits in that trap. 

On one hand, single-qubit gates have a higher fidelity (error rate, $\epsilon$ in range of $10^{-3}$ to $10^{-4}$)~\cite{ionq, qccd-honeywell}. Besides, they can be performed \emph{in-place}. On the other hand, $2$-qubit gates typically have an order of magnitude lower fidelity than $1$ qubit gates ($\epsilon \sim 10^{-2}$)~\cite{ionq, qccd-honeywell}. Fig.~\ref{fig:gates} shows the gate fidelity equation~\cite{murali-ti} for TI systems. This is an experimentally validated gate fidelity model. Here, $\Gamma$ is the trap heating rate, $\tau$ is the gate time, and $\bar{n}$ is the vibrational energy or motional mode of a chain. $A$ is a scaling factor that depends on the number of ions in the chain as $N/ln(N)$. The gate fidelity will degrade if gate time and/or motional mode of the chain increases.
A $2$-qubit gate cannot be applied to ions from different traps. It requires a shuttle.

\subsubsection{Need of shuttle operation:}
Consider the sample program in Fig.~\ref{fig:need-for-shuttle}a. The $4^{th}$ gate in the program \texttt{MS q[a0],q[a1]} involves ions from the same trap (Trap 0) and can be executed in-place or directly (Fig.~\ref{fig:need-for-shuttle}b). However, the $8^{th}$ gate \texttt{MS q[a1],q[a5]} involves ions from different traps (Fig.~\ref{fig:need-for-shuttle}c. Therefore, a shuttle is needed to bring the ions in the same trap. 

\subsubsection{Shuttle steps}
The shuttle operation involves several steps as depicted in Fig.~\ref{fig:Shuttle steps}. First, a$1$ and a$2$ are swapped so that a$1$ is transferred near the shuttle path. Then, a$1$ is split from the chain--$0$ and shuttled/moved from T$0$ to T$1$. The shuttle operation adds energy to the ion. Then, a$1$ is merged to the chain--1. This merge operation increases the vibrational energy ($\bar{n}$) of chain--1. Finally, \texttt{MS q[a1],q[a5]} can be executed as the ions are in the same trap (T$1$) now. As chain--1's $\bar{n}$ is now higher, the subsequent gate operations in this chain (either on aX ions or vX ions) will experience lower fidelity ($F$). \emph{Increasing a chain's motional mode by repeated shuttles is the basis of the attack proposed in this paper}.

\subsection{Initial mapping policy}\label{subsec:im}
The initial mapping - in this context - entails the assignment of program qubits in traps (i.e., mapping of logical qubits to physical traps) and the relative position of qubits inside a trap. For example, the program qubits (a$0$ to a$5$) from the sample program in Fig.~\ref{fig:need-for-shuttle}a are initially mapped as \texttt{T0: [a$0$, a$3$, a$1$, a$2$], T1: [a$4$, a$5$]} (explained in Example~\ref{ex:init-map}). The program execution will start with this allocation, and the mapping will be updated based on shuttles.

The initial mapping policy in ~\cite{murali-ti} is the \emph{greedy} policy where qubits are allocated in the descending order of \emph{edge weights} (wt). A quantum program can be treated as a graph where each node represents a qubit and an edge between two qubits represents a $2$-qubit gate. Thus, the edge weight represents the frequency of a $2$-qubit gate between a pair of qubits. In the Greedy policy, the qubits of the most frequent gates are allocated first.

\begin{example}\label{ex:init-map}
The mapping policy and the result can be explained with the sample program in Fig.~\ref{fig:need-for-shuttle}. The edge weights of the program are as follows: wt(a$0$, a$3$) = $3$ (as the \texttt{MS q[a0],q[a3]} gate appears $3$ times throughout the program), wt(a$1$, a$2$) = $2$, and wt(a$0$, a$1$) = wt(a$4$, a$5$) = wt(a$1$, a$5$) = $1$. Therefore, ions a$0$ and a$3$ are allocated first, then a$1$ and a$2$, and finally, a$4$ and a$5$. In this example, we assume a trap capacity of $4$. Thus, ions a$4$ and a$5$ are in T$1$.
\end{example}

\subsection{Shuttle direction policy}
Shuttle direction policy dictates which ion will be moved to execute a $2$-qubit gate. The shuttle direction policy used in the QCCD compiler~\cite{murali-ti} is illustrated in Listing~\ref{shuttle-policy}. In this paper, we follow the same shuttle direction policy. In the example from Fig.~\ref{fig:Shuttle steps}, both traps have an equal number of ions. Hence, they have the same excess capacity (= trap capacity $-$ \# ions in the trap). Thus, the first ion a1 in the \texttt{MS q[a1],q[a5];} is shuttled from T$0$ to T$1$. 

\begin{lstlisting}[language=Python, caption=Shuttle direction policy~\cite{murali-ti, qccd-github}., label=shuttle-policy]
if excess_cap0 < excess_cap1:
    Move Trap0 --> Trap1
elif excess_cap0 == excess_cap1:
    Move 1st ion of the gate
else:
    Move Trap1 --> Trap0
\end{lstlisting}

\subsection{Multi-programming}\label{subsec:mp}
The proposed attack model exploits the multi-programming setup. We modify the initial mapping policy~\cite{murali-ti, qccd-github} to allow for multi-programming as follows: Suppose, we have two traps (T$0$ and T$1$) and two programs (prog--$0$ and prog--$1$). We allocate prog--$0$ from one end of the T$0$ to T$1$ (if needed) and prog--$1$ from the opposite end of the T$1$ to T$0$. Qubits in a single program are allocated per the greedy mapping policy~\cite{murali-ti}. Allocating multiple programs from opposite ends and different traps ensures that qubits of one program are not mixed with the other program.

\section{Attack Model and Simulation Setup}\label{sec:attack-setup}
\subsection{Attack model}\label{subsec:attack-model}
The following  assumptions are made in the attack model: 
\begin{itemize}

    \item More than one program are running on the QC.
    
    \item The adversary program spans more than one trap, and it shares one trap with another (victim) program.
    
    \item The adversary knows device specification such as trap capacities and communication capacities. This information (especially, trap capacities) are usually public information. For example, the Honeywell $H0$ TI system has a capacity of $6$~\cite{honeywell-h0}.
    
    \item Adversary knows architectural policies: initial mapping policy and shuttle direction policy (can be relaxed. Rationale behind the assumption is discussed in Section~\ref{subsubsec:policy-rational}). 
    
    \item The adversary can access the \emph{compiled} program. This is a reasonable assumption because present quantum clouds provide such access. It allows a user to identify bottlenecks and optimize their programs. For example, in AWS Braket the compiled program is available to the user as the \texttt{MetaData}~\cite{aws-github}.
\end{itemize}

The adversary designs his/her program so that it requires repeated shuttles increasing the vibrational energy of the shared ion-chain and degrading gate fidelities. Note that, the adversary program fidelity takes a hit as a byproduct. However, the objective of the attack is to affect the victim program.

We present two techniques of devising the attack programs: systematic (Section~\ref{sec:systematic}) and random (Section~\ref{sec:random}). On one hand, systematic program generation requires several prior information such as initial mapping and shuttle direction policies. However, it guarantees a linear increase in the number of shuttles (desired) with increased program length. On the other hand, random generation does not guarantee a linear increase of shuttles but requires no prior knowledge. Nevertheless, both methods can degrade victim performance (Section~\ref{sec:results}).

\subsection{Simulation setup}\label{subsec:setup}
In this paper, we use the QCCD compiler-simulator~\cite{qccd-github} accompanying the paper~\cite{murali-ti} to perform simulations. The QCCD compiler takes care of the initial mapping, shuttle insertion, gate scheduling, and fidelity computation of a program. We add our modification on top of the QCCD compiler to allow for multi-programming. Our tweaks include: (i) modifying initial mapping to map multiple programs (as in Section~\ref{subsec:mp}) to the device and (ii) reporting individual program fidelities.

For all simulations, we assume a device with $2$ traps connected in a linear fashion as in Fig.~\ref{fig:attack-intro}a. The trap capacity is $15$ ions per trap with $2$ additional spaces per trap for incoming (shuttled) ions (communication capacity). We also show analysis for trap capacities $20$ and $25$ (with the same communication capacity of $2$). The trap capacity values of $15$ to $25$ are selected as per~\cite{murali-ti} as they observed better performance in this range. 

We set the adversary program size to $18$ qubits (trap capacity $15$ + $3$) so that it spans two traps. We vary victim program sizes from $2$ to maximum remaining space in shared trap T$1$ i.e., $12$ qubits. Note that the adversary and the victim program sizes will change accordingly for trap capacities $20$ ($23$ and $2$ to $17$ respectively) and $25$ ($28$ and $2$ to $22$ respectively). 

$2$-qubit gates mostly affect the program fidelity as they have an order of magnitude lower fidelity than 1-qubit gates. Therefore, we consider only the $2$-qubit gate fidelities without any loss of generality in our analysis. 
The QCCD-simulator~\cite{murali-ti, qccd-github} includes experimentally calibrated parameters~\cite{am2, shuttle-time, qccd-honeywell, am1} for the gate fidelity equation in Fig.~\ref{fig:gates}. The program fidelity is computed from individual gate fidelities ($F$). For a program with $k$ $2$-qubit gates ($g_i$), $i$ being an enumeration parameter across the 2-qubit gates, the program fidelity is $F(g_0) \times F(g_1) \times ... \times F(g_{k-1})$.

\section{Randomized Malicious Program}\label{sec:random}
In this section, we discuss randomized malicious program generation. The randomized attack programs are advantageous as they treat the compiler as a black-box and do not need information about compiler policies. We present a methodology to find and refine an effective random program to launch attacks. The only requirement is that the adversary can submit many programs to the cloud and can access the final compiled program. The final \emph{shuttle-inserted} compiled-program will tell the adversary which program resulted in maximum shuttles. The adversary will pick that program to launch future attacks.

\subsection{General methodologies}
We populate the program with $\binom{18}{2}$ gates (all $2$ qubit combinations from $18$-qubits adversary size with trap cap $15$). Then, we randomly shuffle the gate orders to generate the random program.

We generate $1000$ random circuits and compile them with pseudo-victim programs of sizes from $2$ to $12$ qubits. The idea of a pseudo-victim program is that the adversary will send two programs to the cloud to mimic an adversary-victim pair. After analyzing the collected results, the adversary can select the random circuit that gives the highest average number of shuttles across victim programs of all sizes.

\subsection{Pruning the random circuit}
The best random circuit can be pruned further as not all the gates contribute to shuttling. The intuition is that we can remove some gates from the random circuit without lowering the number of shuttles. The pruning logic is as follows: we remove one gate from the original random circuit starting from the first gate, compile it, and check the number of shuttles. If the number of shuttles does not drop from the original case, we permanently remove the gate from the circuit and move on to the next gate. If removing the gate lowers the number of shuttles, we reinstate the gate and move on to checking the next gate. Following this step-by-step check, we can remove some redundant gates without affecting the number of moves. The pruning on average removed $\approx 48$ gates from the program of $153$ gates.

%%%%%%%%%%%%%%%%%%%%%%%%%%%%%%%%%%%%%%%%%%%%%%%%%%
%%%%%%%%%%% SECTION: SYSTEMATIC Malicious Program
%%%%%%%%%%%%%%%%%%%%%%%%%%%%%%%%%%%%%%%%%%%%%%%%%%

\section{Systematic Malicious Program}\label{sec:systematic}

\subsection{Basic idea}

The systematic method of malicious program generation uses the following $3$ ingredients to craft a strong attack program: (i) initial mapping policy, (ii) shuttle direction policy, and (iii) information on the victim size. As mentioned earlier, a gate will require a shuttle when the ions belong to two different traps. This principle is leveraged in the systematic method, and gates are added in the malicious program with ions from different traps. However, this approach requires knowledge about ion locations, and the above $3$ ingredients facilitate the tracking of ion locations.

%%%%%%%%%%% ALGORITHM 1 %%%%%%%%%%%%
\begin{algorithm}
\SetAlgoLined
\KwIn{trap capacity}
\KwOut{initial mapping controller}
 ion\_list = [$0$ to (trap~capacity $-$ $1$)]\;
 ion\_a, ion\_b $\leftarrow$ $2$ arbitrary ions from ion\_list\;
add gate (ion\_a \& ion\_b) twice in the initial mapping controller block\;
remove ion\_a \& ion\_b from ion\_list\;

\While{ion\_list is not empty}{
    ion\_a $\leftarrow$ ion\_b from last gate\;
    ion\_b $\leftarrow$ next ion from ion\_list\;
    add gate (ion\_a \& ion\_b) twice in the initial mapping controller block\;
    remove ion\_b from ion\_list\;
}
\caption{Create initial mapping controller block.}
\label{algo:im-controller}
\end{algorithm}
%%%%%%%%%%% END ALGORITHM 1 %%%%%%%%%%%%

Our proposed systematic malicious program consists of three blocks: (i) shuttle controller (SC), (ii) a bridging gate, and (iii) initial mapping controller (IMC). Each block is generated using specific logic as explained later in this section. After all blocks are generated, they are stitched to create the complete malicious program (i.e., malicious program = shuttle controller + a bridging gate + initial mapping controller).

The IMC block is generated first, then the SC block, and finally the bridging gate, although they appear in a different order in the program. This ensures no gate from the SC block and the bridging gate have a higher edge weight than gates from the IMC block (explained more in Section~\ref{subsubsec:im-explanation}). 

\subsection{Initial mapping controller}
With knowledge about the initial mapping policy, the adversary can intelligently add gates in the program to force a known initial mapping. As described in Section~\ref{subsec:im}, the initial mapping policy in the QCCD-compiler is a greedy one that allocates ions based on gate frequencies (edge weights). Therefore, the adversary can judiciously increase edge weights between certain nodes (ions/qubits) which he/she wants to be allocated first. Algorithm~\ref{algo:im-controller} illustrates the gate selection methodology. We explain the algorithm with Example~\ref{ex:init-controller}.

\begin{figure}[t]
    \centering
    \includegraphics[width=2.7in]{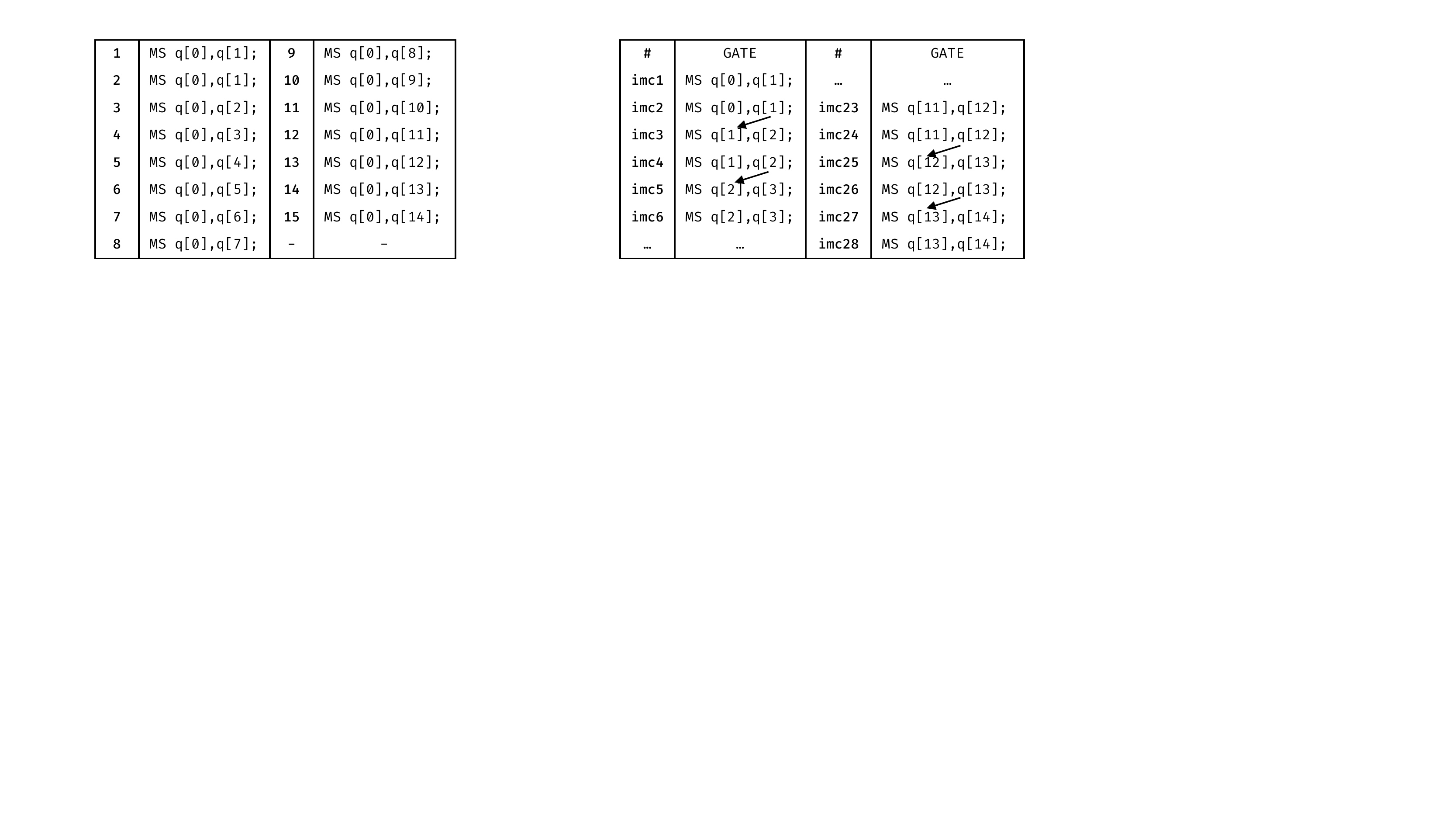}
    \caption{IMC block of the malicious program. Arrows showing dependency between gates.\vspace{-5mm}}
    \label{fig:preamble}
\end{figure}
%%%%
\begin{example}\label{ex:init-controller}
Suppose, the trap capacity is $15$. Thus, the ion\_list will be [$0$, $1$, $2$, ..., $13$, $14$]. Next, we arbitrarily select two ions $0$ (ion\_a) and $1$ (ion\_b) from the list. Note that any two ions can be selected. We add gate \texttt{MS q[0], q[1]} twice in the program with these ions. This will make edge weight of ($0$, $1$) = $2$. Next, ions $0$ and $1$ are removed from the ion\_list which now becomes [$2$, $3$, $4$..., $13$, $14$]. 

Next, ion\_a and ion\_b values are updated. ion\_a's value becomes the previous ion\_b value (i.e., $1$), and ion\_b's value becomes the next value from the ion\_list (i.e., ion\_b = $2$). Then, the gate with these two ions - \texttt{MS q[1], q[2]}- are added twice in the block. Finally, ion\_b = $2$ is removed from the ion\_list for this iteration, and ion\_list becomes [$3$, $4$, $5$, ..., $13$, $14$].
The above routine is repeated unless the ion\_list becomes empty i.e., all the ions are added in the block.

The final IMC block will be similar to Fig.~\ref{fig:preamble}. All the gates in the block have an edge weight of $2$.  
The logic in the other blocks (SC block and bridging gate) ensures that no other edge weight exceeds $1$ (i.e., all other gates will appear once).
Due to the higher edge weights, gates in the IMC block will be allocated first according to the greedy policy. Thus, ions $0$ to $14$ ($15$ qubits) will be allocated first to T$0$ (the remaining $3$ qubits of the $18$ qubit adversary will be allocated to T$1$ by default).
\end{example}

Assuming a victim size of $12$, the trap states after initial mapping will be \texttt{\{T0 (EC = 2): [0, 1, 2, ..., 13, 14], T1 (EC = 2): [15, 16, 17] + [12Q victim]\}}. Here, EC = excess capacity. T$0$ has $15$ ions from adversary program. T$1$ also has $15$ ions, $3$ from the adversary and $12$ from the victim. Thus, each trap has an excess capacity of $2$ (from communication capacity).

\subsection{Shuttle controller}
After the IMC block ensures a known initial mapping, we use Algorithm~\ref{algo:shuttle-gates} to add gates in the malicious program that require shuttles. The flow consists of $5$ steps. We explain each step in Example~\ref{ex:shuttle-gates}. The core idea is to track ion locations (using shuttle direction policy) after each gate and select ions from different traps for the next gate.

\begin{example}\label{ex:shuttle-gates}

\begin{algorithm}
\SetAlgoLined
\KwIn{trap states, shuttle direction policy, node weights, edge weights, block length, prog. size}
\KwOut{shuttle controller block (sc\_block)}
 sc\_block $\leftarrow$ empty;  
 \# of added gates $\leftarrow$ 0; 
 flag $\leftarrow$ 0\;
 \While{\# of added gates $<$ block length}{
  \textcolor{mybrown}{// STEP -- 1\;}
  \uIf{\# of added gates == $0$}{
   $ion\_a$ $\leftarrow$ random ion $\in$ \{0 to prog. size$-1$\}\;
   }
  \uElseIf{flag == 0} 
   {
   $ion\_a$ $\leftarrow$ moved ion from last gate\;
  }
  \Else{
    $ion\_a$ $\leftarrow$ non-moved ion from last gate\;
  }
  \textcolor{mybrown}{// STEP -- 2\;}
   get ion\_a's location; 
   get opposite\_trap\; % $\leftarrow$ \{T$0$, T$1$\} - ion\_a's trap\;
  
  \textcolor{mybrown}{// STEP -- 3\;}
  \For{ion\_b $\in$ \{ions in the opposite trap\}}{
    \eIf{edge weight (ion\_a, ion\_b) == $0$}{
        flag $\leftarrow$ 0; 
        break\;
    }{
        flag $\leftarrow$ 1; 
        continue\;
    }
  }
  \If{flag == $1$}{
    continue\;
  }
  
  \textcolor{mybrown}{// STEP -- 4\;}
  sc\_block $\leftarrow$ sc\_block + gate (q[ion\_a], q[ion\_b])\;
  last\_gate $\leftarrow$ (ion\_a, ion\_b)\;
  \textcolor{mybrown}{// STEP -- 5\;}
  identify moved ion, update node and edge weights, update trap states using shuttle policy\;
  
  \# of added gates $\leftarrow$ \# of added gates$+ 1$
 }
 \caption{Create shuttle controller block \vspace{-5mm}}
 \label{algo:shuttle-gates}
\end{algorithm}

\begin{figure}
    \centering
    \includegraphics[width=2.5in]{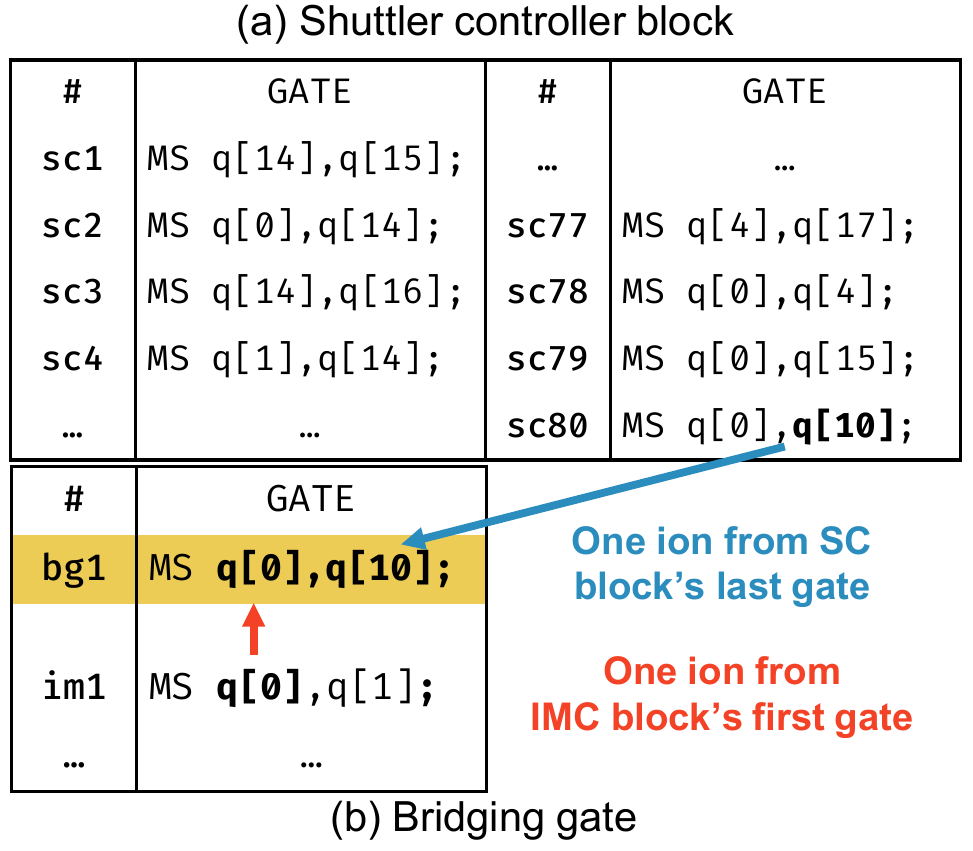}
    \caption{(a) The partial SC block, showing first and last few gates. (b) The bridging gate between the shuttle controller and the initial mapping controller block. Only one bridging gate is necessary between these blocks. \vspace{-4mm}}
    \label{fig:scb}
\end{figure}

\textbf{Step--1:} For the very first gate, we randomly select $1$ ion from adversary's ion list (i.e., from $[0$ to $17]$ for the $18$-qubit adversary). Suppose, the selected ion is $14$ (i.e., ion\_a = $14$). 

\textbf{Step--2:} From the initial mapping, we know ion $14$' is in T$0$. Therefore, the opposite trap is T$1$..

\textbf{Step--3:} Select ion $15$ from T$1$. Check if gate \texttt{MS q[14], q[15]} does not exist in the program (i.e., edge weight of ($14$, $15$) is $0$ in the program graph). As the gate does not exist in the program, we do not check more ions from T$1$ and can break from the loop with $ion\_b = 15$. 

\textbf{Step--4:} Add gate \texttt{MS q[14], q[15]} in the program. This gate will require a shuttle when executed. Therefore, trap states need updating after the gate.

\textbf{Step--5:} Both T$0$ and T$1$ have equal excess capacity of $2$ each. Therefore, according to the shuttle direction policy in Listing~\ref{shuttle-policy}, the first ion in the gate \texttt{MS q[14], q[15]} i.e., ion $14$ will move from T$0$ to T$1$. Thus, the updated trap states are \texttt{\{T0 (EC = \textbf{3}): [0, 1, 2, ..., 13], T1 (EC = \textbf{1}): [\textbf{14}, 15, 16, 17] + ([12Q victim]\}}.
(Note, the victim size information is required to compute the excess capacity of the shared trap T$1$ and to find the shuttle direction accurately.) Edge weights list is updated with the new gate.

Following the same routine, we keep adding gates in the malicious program until target number of gates are reached. For the next iteration, we pick ion $14$ (moved ion from the last gate) as the $ion\_a$ in Step--1. As ion $14$ is in T$1$ now, we pick the other ion from T$0$ (say, ion $2$). The next gate is \texttt{MS q[2], q[14]}. As T$0$ has more EC (= $3$) than T$1$ (EC = $1$), ion $14$ will again move but this time from T$1$ to T$0$. Finally, the updated trap states after this gate will be \texttt{\{T0 (EC = \textbf{2}): [0, 1, 2, ..., 13, \textbf{14}], T1 (EC = \textbf{2}): [15, 16, 17] + [12Q victim]\}}. A partial shuttle controller block of block length $80$ is illustrated in Fig.~\ref{fig:scb}a. Note the last gate in the block. It is required for the bridging gate.
\end{example}

\begin{figure}[b]
    \centering
    \includegraphics[width=2.8in]{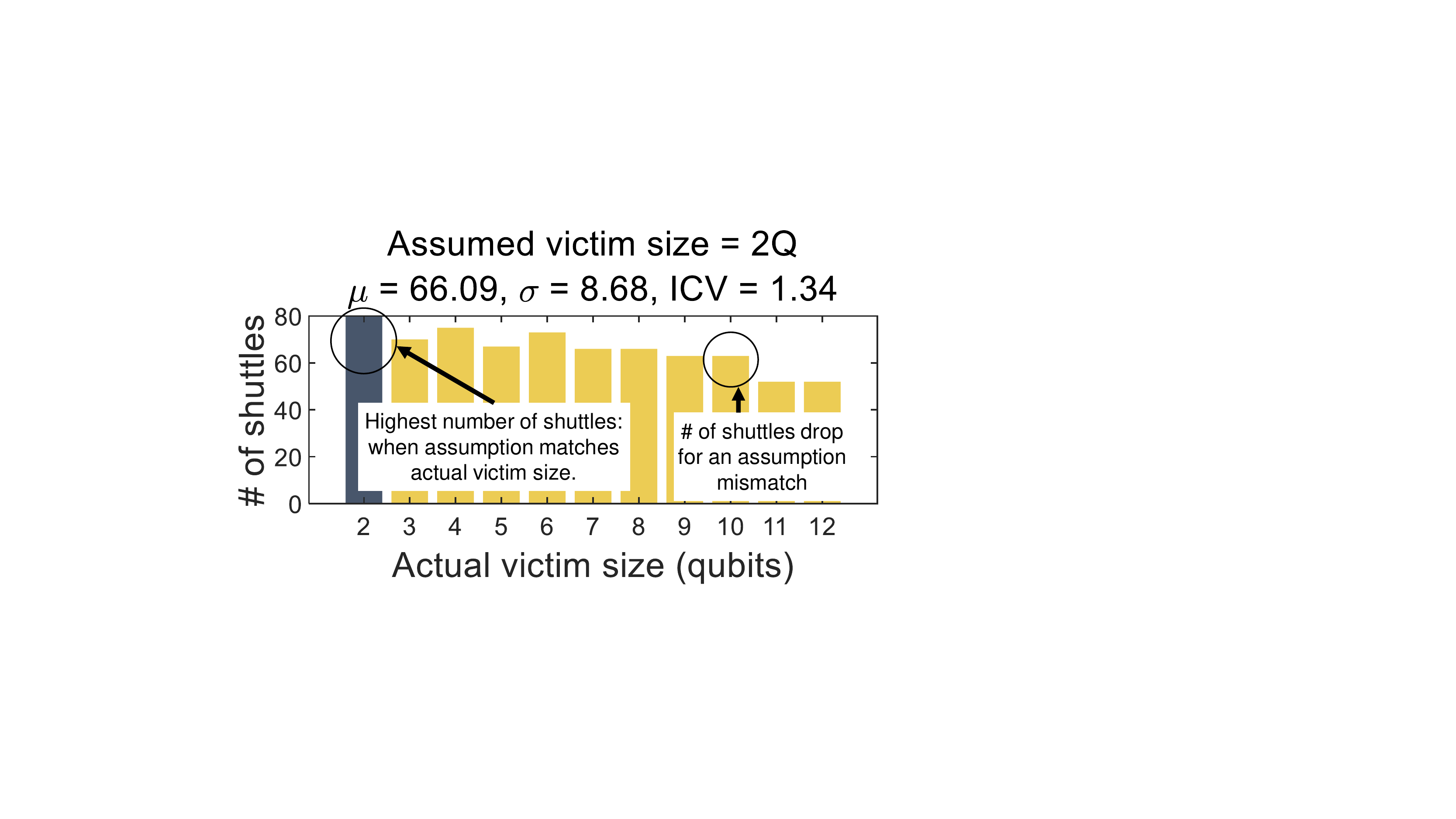}
    \caption{Variation of shuttle numbers across different (actual) victim sizes. Here, the adversary program is designed assuming a victim size of $2$. \vspace{-4mm}}
    \label{fig:num-moves-2q}
\end{figure}

\begin{figure*}
    \centering
    \includegraphics[width=1.0\linewidth]{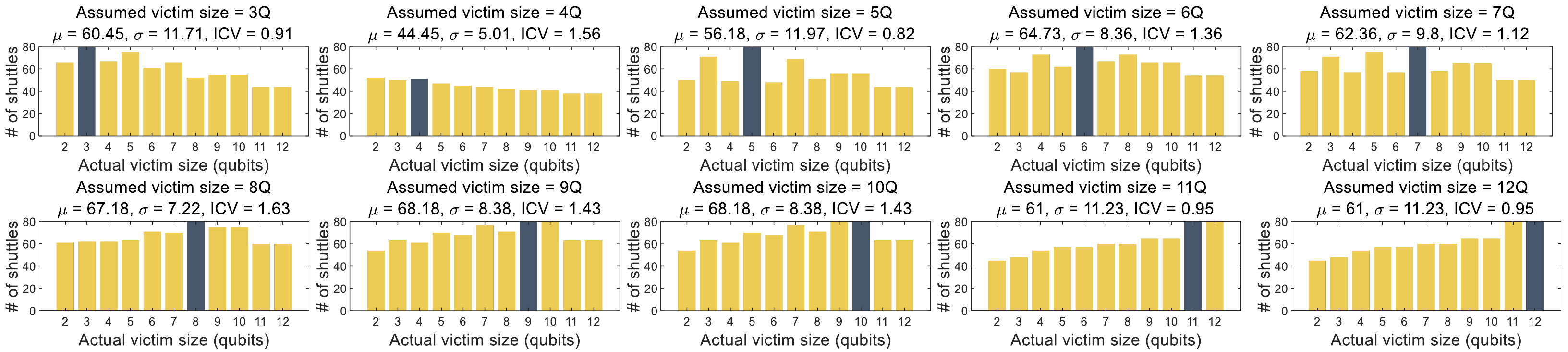}
    \caption{Effect of victim size assumption (trap capacity $=$ $15$). Assuming victim size $8$ provides best result across all actual victim sizes (highest ICV). \vspace{-4mm}}
    \label{fig:var-num-moves}
\end{figure*}

\begin{figure}
    \centering
    \includegraphics[width=2.8in]{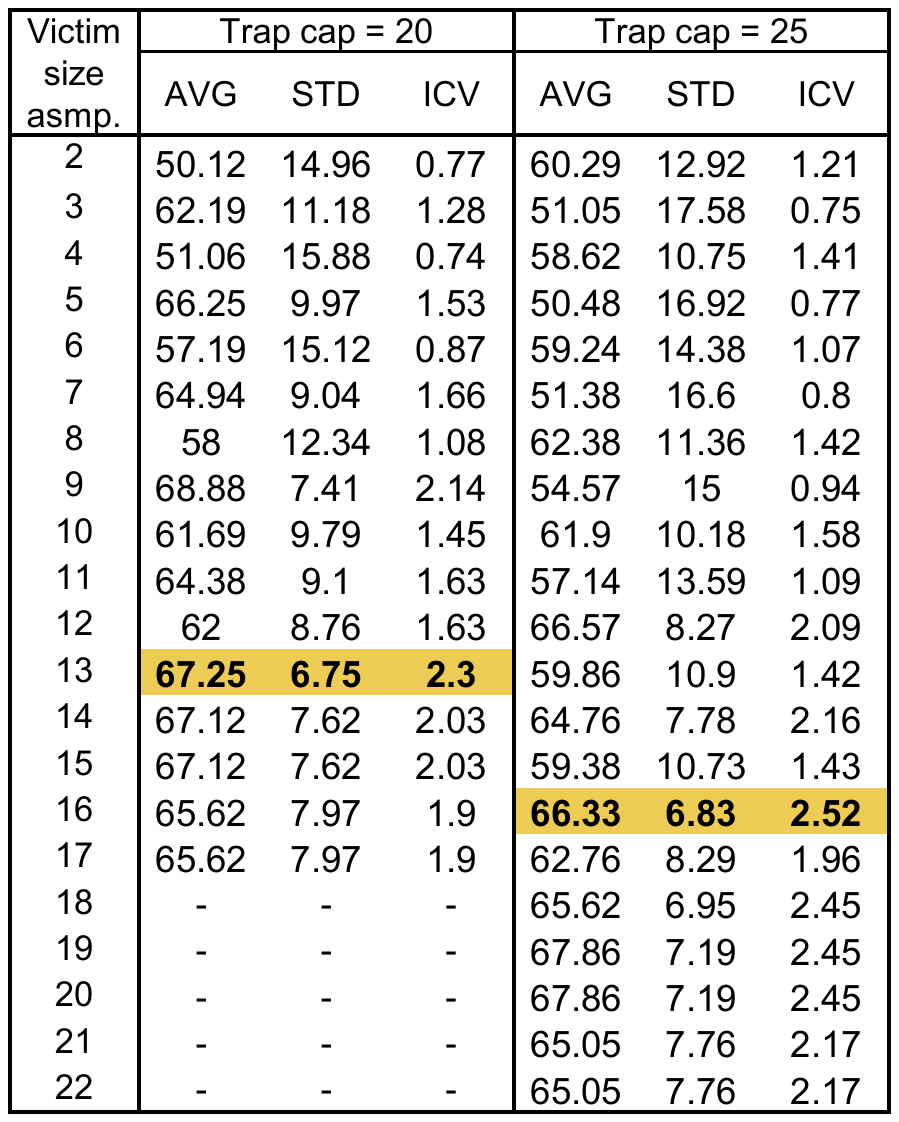}
    \captionof{table}{Statistics from victim size-assumption sweep for trap caps $20$ and $25$. For trap cap $20$, assuming a victim size of $13$ will provide best result (high ICV) across all actual victim sizes, and for trap cap $25$ the best assumption is $16$. \vspace{-3mm}}
    \label{tab:num-moves-20-25}
\end{figure}

\subsection{The bridging gate and the complete malicious program}
\subsubsection{The bridging gate}
The bridging gate is formed by taking one ion from the \emph{last gate} of the shuttle controller block and one ion from the \emph{first gate} of the initial mapping controller block (Fig.~\ref{fig:scb}b). As the name suggests, this gate bridges two blocks and maintains the gate dependency. Only one bridging gate is necessary. 

\subsubsection{The complete malicious program}
The complete malicious program is created by combining individual parts in the following order: gates from the SC block + bridging gate + gates from the IMC block. 

\subsection{Discussions on the systematic method}
\subsubsection{Out-of-order IMC block}\label{subsubsec:im-explanation}
\emph{In greedy policy, ions from a gate is primarily mapped based on their frequency of appearance and secondarily by their order of appearance in the program.} Gates in the IMC block have the highest frequencies across the complete program because they each are deliberately added twice in the program, and gates from the other blocks each are added once. Generating the IMC block first ensures that other blocks can skip gates already in the IMC block. The compiler sorts gates in the descending order of their frequencies for mapping. Therefore, gates in the IMC block come first in the sorted list for mapping although they appear last in the malicious program.

\subsubsection{Necessity of victim size information}
The victim size is a parameter of the SC block generation algorithm. If the victim size is correct, each gate in the generated SC block will require one shuttle when the malicious program is executed. However, with an inaccurate victim size, some of the gates in the SC block will not force a shuttle. 
\begin{example}
Consider the SC block in Fig.~\ref{fig:scb}a generated \emph{assuming} a victim size of $12$. Suppose, the actual victim size during run is $5$. Then, the actual trap states and excess capacities will be as follows at the beginning: \texttt{T0 (EC = 2): [0, 1, ..., 14], T1 (EC = 7): [15, 16, 17] + [5Q victim]}. Gate \# sc$1$ will need a shuttle as ion $14$ is in T$0$ and ion $15$ is in T$1$. Updated trap states will be \texttt{T0 (EC = \textbf{3}): [0, 1, ..., 13], T1 (EC = \textbf{6}): [\textbf{14}, 15, 16, 17] + [5Q victim]} (Note T$1$ EC $>$ T$0$ EC, different than the assumption). Gate \# sc2 will also require a shuttle as the ions are in different traps. New trap states: \texttt{T0 (EC = \textbf{4}): [1, ..., 13], T1 (EC = \textbf{5}): [\textbf{0}, 14, 15, 16, 17] + [5Q victim]}. However, the next gate \# sc3 will not require a shuttle as both ions $14$ and $16$ are in T$1$.
\end{example}

The above example illustrates that some gates in the SC block will skip shuttling when the actual victim size is different from the design assumption. As the adversary may not know the victim size, he/she needs to assume a value that gives the best number of shuttles across all possible victim sizes. To find the best assumption, we sweep the victim size and propose selecting the value that gives the highest inverse coefficient of variation (ICV). The inverse of coefficient of variation is the ratio of the mean ($\mu$) to standard deviation ($\sigma$). 
A higher ICV indicates a higher mean and/or lower standard deviation. For our purpose, we want a distribution that provides a higher mean number of moves with a tighter spread (a lower $\sigma$). Before computing ICV, we normalize both $\mu$ and $\sigma$ with respect to the respective maximum value. The results are discussed in Section~\ref{sec:results}.

\subsubsection{Availability of architectural policies}\label{subsubsec:policy-rational}
The systematic method relies on knowledge initial mapping and shuttle direction policies to ensure an effective attack. Details on such architectural policies are available in literature~\cite{poulami-mp,murali-noise-adaptive, tannu, siraichi2018qubit} and/or compiler documentations~\cite{rigetti-compiler, qiskit-transpiler}. 
Architectural policies like speculative execution are available to the public even in the classical domain. Numerous papers on architectural policy flaws and their mitigation also exist~\cite{spec-exec-attk-1, spec-exec-attk-2, spctre, meltdown}. \emph{Spectre}~\cite{spctre} and \emph{Meltdown}~\cite{meltdown} are two such famous vulnerabilities that exploit the knowledge of an architectural policy. Thus, we envision such availability of architectural policies in the quantum domain as well. Finally, if the cloud provider does a good job in protecting their secrets i.e., restrict access to policies, the adversary can follow two options. \emph{Option--1:} the adversary can adopt a trial-and-error route and design several attack programs each considering separate policies available in the literature. Then, he/she can launch an attack to check which one gives the best results (high number of shuttles). \emph{Option--2:} the adversary can revert to the random attack described in Section~\ref{sec:random} which does not require any prior information.

\section{Results and Discussions}\label{sec:results}

\subsection{Victim size sweep}
\subsubsection{Systematic attack}
Fig.~\ref{fig:num-moves-2q} shows the number of shuttles for various actual victim sizes ($2$ to $12$, trap cap $15$) where the attack program is designed assuming a $2$-qubit victim. The plot shows the highest number of moves ($80$) is achieved when adversary assumption matches actual victim size (dark bar in the plot). We also observe that for other actual victim sizes shuttle numbers vary and typically drop from the highest value. We report the mean ($\mu$), standard deviation ($\sigma$), and inverse of coefficient of variation (ICV) of this distribution. 

To find the best assumption, we sweep the victim size assumption from $3$ to $12$ (in addition to the assumed size of $2$ in Fig.~\ref{fig:num-moves-2q}) and record the statistics. The results are plotted in Fig.~\ref{fig:var-num-moves}. In all cases, we observe that the highest number of moves is achieved when the assumption matches the actual size. From the statistics, an assumed victim size of $8$ qubits gives the best performance (highest ICV). 

We perform the same analysis for trap capacities (cap) $20$ and $25$. For trap cap of $20$, possible victim sizes vary from $2$ to $17$, and for trap cap $25$ it varies from $2$ to $22$. We report the mean, standard deviation, and ICV for both trap caps in Table~\ref{tab:num-moves-20-25}. For trap cap $20$, an assumption of $13$ qubits in the victim gives the highest ICV, and for trap cap $25$ the best assumption is $16$ qubits.

\subsubsection{Random attack}
\begin{figure}
    \centering
    \includegraphics[width=2.8in]{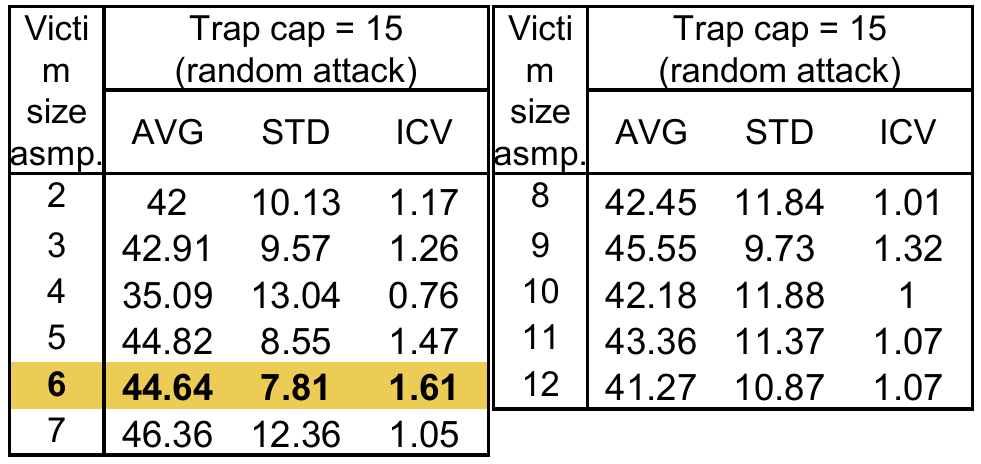}
    \captionof{table}{Shuttle statistics for the random attack. \vspace{-4mm}}
    \label{tab:table-rand-15}
\end{figure}

Table~\ref{tab:table-rand-15} shows the statistics for victim size sweep for random attack programs. The assumption of victim size $6$ provides the best results in terms of ICV. Therefore, we select this program to launch future attacks.

\begin{figure}
    \centering
    \includegraphics[width=2.8in]{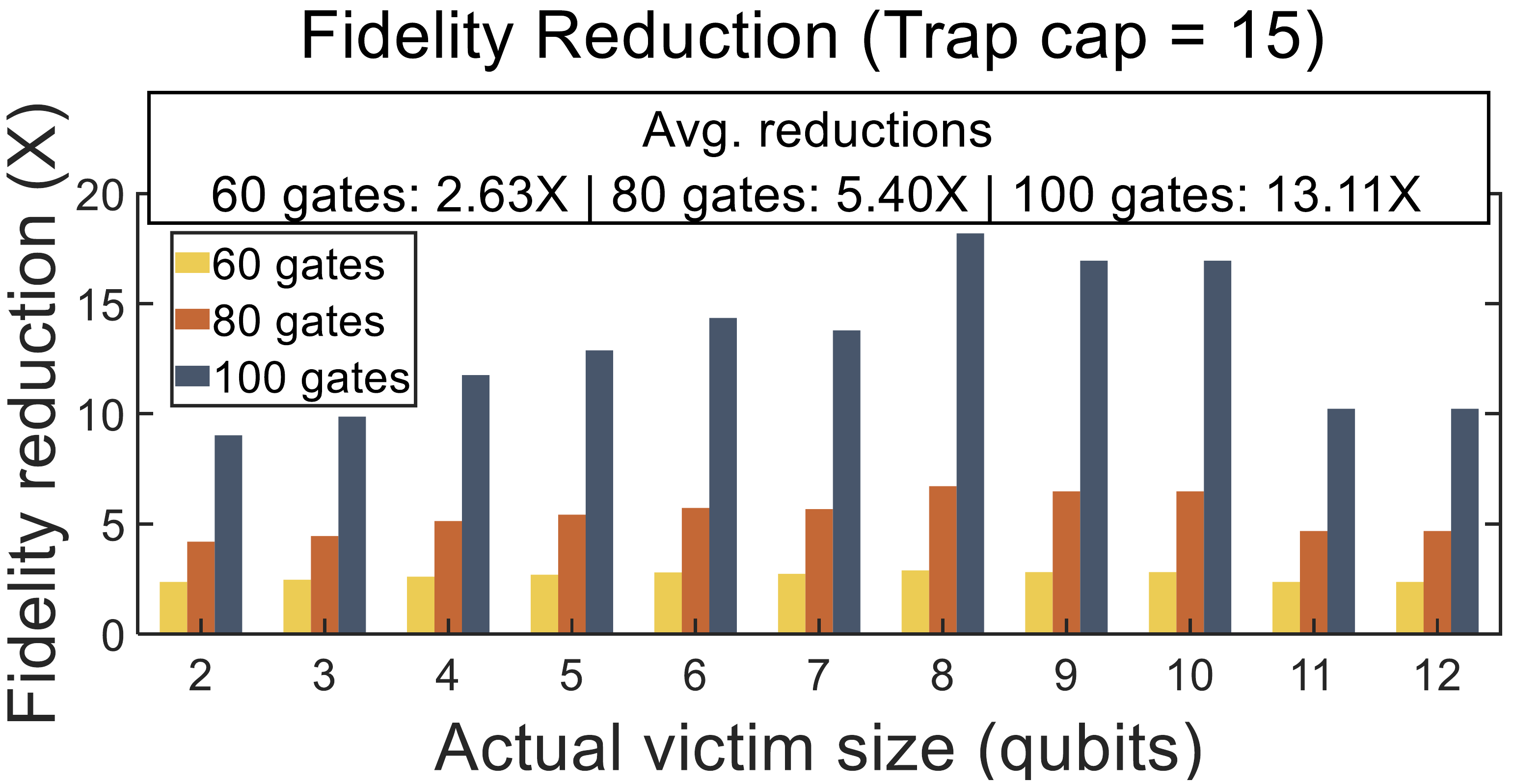}
    \caption{Fidelity reduction for various victim lengths. The adversary program assumes victim size of $8$. The trend exhibits a positive correlation between reduction and length. \vspace{-7mm}}
    \label{fig:fid-reduction}
\end{figure}

\subsection{Fidelity reduction}
\subsubsection{Systematic attack}
\begin{figure}
    \centering
    \includegraphics[width=2.6in]{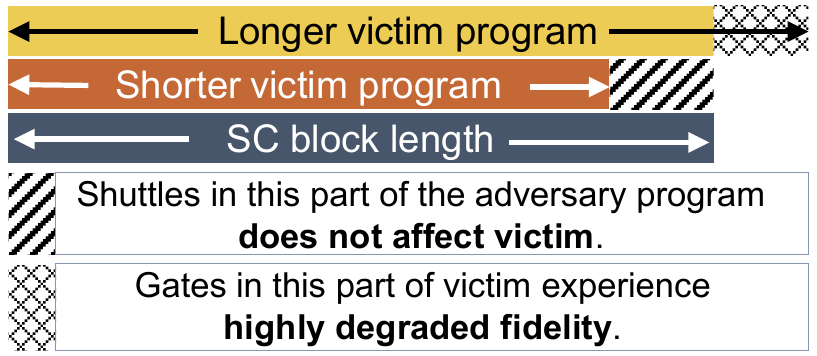}
    \caption{Explanation of the positive correlation between fidelity reductions and victim lengths.}
    \label{fig:fid-reduction-explanation}
\end{figure}

Fig.~\ref{fig:fid-reduction} shows fidelity reduction of the victim program under attack for $3$ victim lengths - $60$, $80$, and $100$. The SC block length, trap cap, and assumed victim size are $80$ gates, $15$ ions, and $8$ qubits respectively. The plot shows that the fidelity reduction increases with higher victim lengths. Average fidelity reductions are $2.63$X, $5.40$X, and $13.11$X for victim lengths $60$, $80$, and $100$ respectively. Fig.~\ref{fig:fid-reduction-explanation} qualitatively explains the positive correlation between victim length and fidelity reduction. The shuttle operations are spread across the length of the SC block. If the victim program completes before the SC block, shuttles at the later part of the block (hatched pattern in Fig.~\ref{fig:fid-reduction-explanation}) does not affect the victim program. 
The opposite happens if the victim program is longer. Repeated shuttles from adversary programs increase the chain energy. Therefore, the later gates (cross pattern in Fig.~\ref{fig:fid-reduction-explanation}) in the victim program experience excessive fidelity drops. 

\begin{figure}
    \centering
    \includegraphics[width=2.5in]{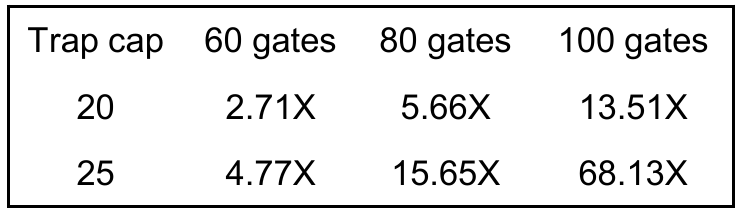}
    \captionof{table}{Fidelity reduction values for trap capacities $20$ and $25$. Results for trap capacity $20$ experiences slightly elevated reductions than $15$, and for trap capacity $25$, they are even worse. This can be attributed to scaling factor $A$ of the motional mode ($\bar{n}$) (Fig.~\ref{fig:gates}). With more ions in the trap, $A$ increases leading to more pronounced effect of $\bar{n}$. \vspace{-4mm}}
    \label{fig:fid-red-20-25}
\end{figure}

Table.~\ref{fig:fid-red-20-25} shows fidelity reductions for trap caps $20$ and $25$ for $3$ victim program lengths - $60$, $80$, and $100$. For trap cap $20$, the average fidelity reduction for these $3$ program lengths are $2.71$X, $5.66$X, and $13.51$X. For trap cap $25$, the average fidelity reductions are $4.77$X, $15.65$X, and $68.13$X. In case of trap cap $25$, we observe aggravated fidelities, especially for larger victim sizes. 
Intuitively, this behavior can be attributed to the scaling factor $A$ in the gate fidelity equation $F = 1 - \Gamma \tau - A(2\bar{n}  +1)$. Factor $A \propto N/log(N)$ where $N$ = number of qubits in the chain. For larger victims in a larger capacity trap, $A$ scales up making the motional mode ($\bar{n}$) more pronounced, and exacerbating gate fidelity ($F$).

\subsubsection{Random attack}
\begin{figure}
    \centering
    \includegraphics[width=3in]{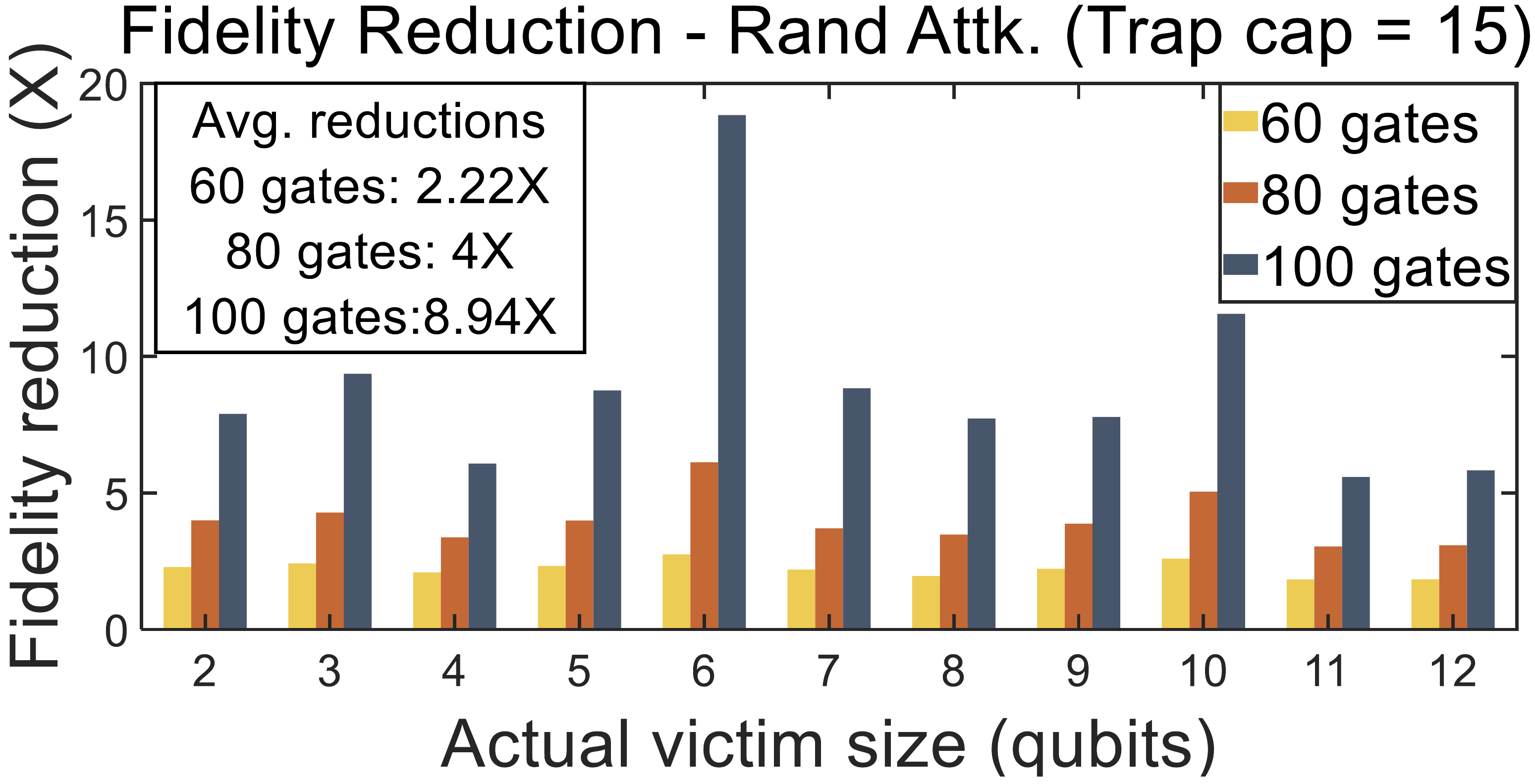}
    \caption{Fidelity reduction for random attack program. \vspace{-5mm}}
    \label{fig:fid-red-rand}
\end{figure}

Fig.~\ref{fig:fid-red-rand} shows the fidelity reduction values for random attack program (trap cap = $15$). We omit the values for trap cap $20$ and $25$ for brevity (fidelity reductions will be even higher at these capacities). The results show an average fidelity reduction of $2.22$X, $4.0$X, and $8.94$X. These values are lower than the systematic attack.

\subsection{Choice between systematic and random attack programs}
The choice between systematic and random attack is not an either-or proposition although the systematic approach provides a higher fidelity reduction. The adversary needs to submit many programs to the cloud to find a good attack program using the random approach. Using the systematic approach, a good attack program can be generated in one try. Thus, the choice between approaches will depend on the resources available to the adversary. If he/she has information about the architectural policies, adopting the systematic approach is the fastest and the best choice. If an adversary has the resources to run many programs on the cloud (running programs will cost money) and/or does not have knowledge about necessary policies, then adopting the random approach will lead to stronger attacks.

\section{Countermeasures}\label{sec:countermeasures}
\subsection{Random initial mapping}
The compiler can adopt a random initial mapping policy where each program at each iteration starts from a random allocation. Both malicious program creation methods rely on a consistent initial mapping i.e., the same program will be allocated in the same fashion for every run. In case of a random attack, if initial mapping changes randomly from one instance to another, the attack program generated at one instance (with one mapping) will not work effectively for another instance (with a different) mapping. A random mapping will invalidate the systematic method as traps will start from unknown states.

We validate this proposed technique with an $18$-qubit adversary program designed for a $12$-qubit victim. With the greedy policy, it forces $80$ shuttles. Next, we switch the initial mapping policy to random and gather results for $1000$ runs. We find that with a random initial mapping policy, the average number of moves drastically drops to $\approx 27~(\sigma = 9.45)$ ($2.96$X drop). Therefore, it proves the efficacy of random initial mapping in weakening the attack.

\subsubsection{Trade-off of random initial mapping}
\begin{figure}
    \centering
    \includegraphics[width=3in]{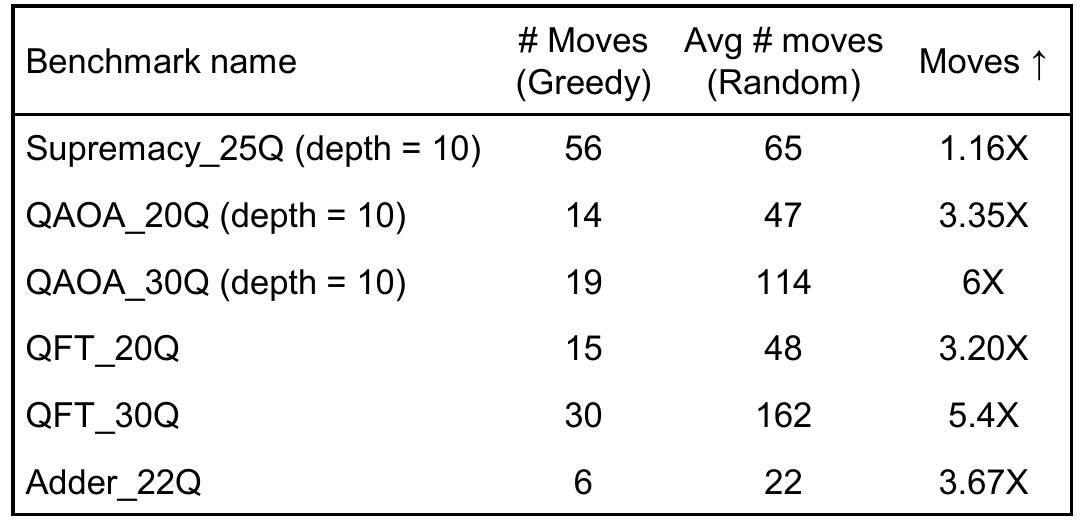}
    \captionof{table}{Penalty of random mapping in NISQ benchmarks. \vspace{-8mm}}
    \label{tab:rand-map-penalty}
\end{figure}

Although a random initial mapping policy can disarm an adversary, it may penalize a legitimate user. A good initial mapping policy tries to place ions with frequent gates together so that communication can be minimized in typical benchmarks (note that attack programs are not typical programs-- they are artificially crafted to contain numerous shuttles by hacking an intelligent initial mapping policy). However, a random initial mapping policy does not exploit such intelligence and cannot always guarantee an optimal number of communications. To show the impact of random mapping, we simulate a suite of popular noisy intermediate-scale quantum (NISQ) benchmarks used in~\cite{murali-ti} with both greedy and random mapping policies. Benchmarks include quantum Fourier transform (QFT), quantum approximation optimization algorithm (QAOA) circuit, supremacy circuit from Google's quantum supremacy experiment, and quantum adder circuit. The circuits are generated using~\cite{qcg}. The mean values for random policy are computed from $1000$ random allocations. The results (Table~\ref{tab:rand-map-penalty}) show that random mapping increases shuttles for NISQ benchmarks up to $6$X. Thus, a random mapping policy can weaken attacks at a cost of penalizing legitimate users.

\subsubsection{Hybrid approach}
To alleviate the issue, the cloud can adopt a hybrid approach. In the hybrid approach, the compiler compiles a program with both random and intelligent (e.g., greedy) initial mapping and discards the result with higher shuttles. It will execute the version with the lower shuttles. 

\subsection{Dummy pad qubits in the victim program}
A user (victim) can protect his/her program by adding a sufficient number of dummy qubits to pad the unused qubits in a trap. Suppose, the actual user program needs $10$ qubits, and he/she wants to execute the program on a system with a trap cap $15$/trap. In such a case, the user can add $5$ dummy pad qubits in his/her program to make the program size $15$ which will fully occupy a trap. Thus, an adversary qubit cannot share a trap with the victim preventing shuttle-induced fidelity degradation. The user can apply virtual-Z gates~\cite{virtual-z} (e.g., GZ gate in the IonQ machine~\cite{ionq-gates}) on the dummy qubits. It will ensure that the compiler considers the qubits during allocation. As the virtual-Z gate is a software gate, it has perfect fidelity, requires no physical, and will not affect the user program. 

However, there can be a security vs. cost trade-off. The quantum cloud may charge the user more for using more qubits. 
Consider a linear cost model where requesting $1$ qubit cost $1$ unit. The cost of running the $10$-qubit will be $10$ units. However, with $5$ dummy qubits for padding and security, now the user has to spend $15$ units increasing the cost by $1.5$X. Not to mention that the cost model could be based on an exponential relation with the qubits counts. Therefore, this defense will be more cost-effective for a low number of dummy qubits (i.e., where actual user program size is large and/or trap cap is low).

\subsection{Capping maximum number of allowed shuttles}
The cloud can enforce a \emph{max shuttle} to prevent shuttle-exploiting attacks. The cloud provider can check the required number of shuttles in a program, and if it exceeds the set maximum value, the cloud can schedule it separately (without any accompanying program). This means for certain programs the cloud will dynamically switch to a single-programming mode from the multi-programming mode. However, the cloud will lose some throughput due to this switching. It can cover the loss by charging extra for programs requiring a high number of shuttles. Suppose, the $18$ qubit adversary program exceeds the max set shuttles. The cloud needs to run it in a single-programming mode with $30$ qubit resources ($2$ traps $\times 15$ qubits/trap). Thus, the cloud will now charge $30$ units - considering the previous linear cost model - instead of $18$ units before to cover the loss in device utilization. In this way, programs with high shuttles cannot affect other programs, and the cloud will not incur a loss.

\section{Conclusion}\label{sec:conclusion}
In this paper, we present a vulnerability in multi-programming access to TI quantum computers and propose several defenses. The key takeaway message of the paper is to establish shuttle operations as a mode of attack. We also present several mitigation measures such as, hybrid mapping policy, padding user programs with dummy qubit, and setting a maximum allowed number of shuttles.  

\section*{Acknowledgement}
This work is supported by National Science Foundation (NSF) (OIA-$2040667$ and DGE-$2113839$) and seed grants from Penn State Institute for Computational and Data Sciences (ICDS) and Penn State Huck Institute of the Life Sciences.

% \newpage
\bibliographystyle{./bibliography/IEEEtran}
\bibliography{./bibliography/IEEEabrv,./bibliography/ref}

\end{document}